\documentclass{PoS}
\usepackage{mcite}
\usepackage{amsmath}
\usepackage{amsfonts}
\usepackage{amssymb}
\usepackage{mathrsfs}
\title{$\bf m_c/\bf m_s$ with Brillouin improved Wilson fermions}
\ShortTitle{$m_c/m_s$ with Brillouin fermions}

\author{Stephan D\"urr\\
        J\"ulich Supercomputing Center at Forschungszentrum J\"ulich and Wuppertal University\\
        E-mail: \email{s.durr@fz-juelich.de}}

\author{\speaker{Giannis Koutsou}\\
        Computational-based Science and Technology Research Center\\
        E-mail: \email{g.koutsou@cyi.ac.cy}}

\abstract{We present a calculation of the ratio of the charm quark mass to the strange quark mass. Using the Brillouin improved Wilson action, we are able to calculate this ratio in a single framework, using a relativistic fermionic action throughout. The calculation is carried out on selected ensembles of two flavor clover improved lattices produced by the QCDSF collaboration, allowing an extrapolation to the continuum, to infinite volume and to the physical pion mass.}

\FullConference{ The XXIX International Symposium on Lattice Field Theory - Lattice 2011\\
July 10-16, 2011\\
Squaw Valley, Lake Tahoe, California}

\begin{document}

\section{Introduction}
In this presentation we detail our measurement of the charm to strange quark mass ratio $m_c/m_s$~\cite{Durr:2011ed}. The exact determination of the masses of quarks is of particular phenomenological interest, since they cannot be measured directly in experiments, leading to rather large errors in their PDG values~\cite{Nakamura:2010zzi}. This is an area where lattice QCD can provide important input, especially in the case of ratios of quark masses where lattice renormalization factors cancel, which otherwise require dedicated renormalization techniques to keep the systematic error small.

We use the Brillouin action, which we introduced in Ref.~\cite{Durr:2010ch}, to calculate $m_c/m_s$. Among a number of nice features, the Brillouin action has been shown to exhibit small cut-off effects even in the charm quark mass region when compared to regular clover-improved Wilson. In the sea sector, we select a set of 2-flavor clover-improved lattice gauge ensembles provided by the QCDSF collaboration which allow us to extrapolate to zero lattice spacing, to the physical pion mass and to infinite volume. We compare our result, and find agreement to the two recent state-of-the-art calculations of this quantity by the HPQCD collaboration~\cite{davies:2009ih} and the  ETM collaboration~\cite{blossier:2010cr}.

\section{Lattice Setup}
In this work we use a mixed action approach to determine the ratio $m_c/m_s$. In the sea sector, we use configurations generated by the QCDSF collaboration. The parameters of the gauge configuration ensembles used are detailed in Table~\ref{table:configs}, while for the simulation details see 
Refs.~\mcite{
Bietenholz:2010az, Collins:2011mk}. The selection of the specific subset of configurations was chosen such as to provide a wide enough window in the pion mass, the physical volume and the lattice spacing in order to allow for an extrapolation to the physical point. In the valence sector of our mixed action approach we use our recently developed Brillouin improved Wilson action~\cite{Durr:2010ch}. As in our original publication of this fermion action, we perform a single step of APE-smearing on the gauge links to further improve gauge-noise and the approach to the continuum, as well as tree-level clover improvement ($c_{SW} = 1$). 

As will be more apparent in the following sections, in our approach we need only set the scale in the final step, when performing the extrapolation to the physical point. In that step, the scales used are taken from Ref.~\cite{Collins:2011mk}, using the Sommer scale as $r_0$ = 0.5~fm at the chiral limit. This yields $a$ = 0.076, 0.072 and 0.060 fm for $\beta$ = 5.25, 5.29 and 5.4 respectively.


\begin{table}[!h]
  \caption{Table of the QCDSF configurations used. For each ensemble $\sim$500 configurations were used.}
  \label{table:configs}
  \begin{center}
    \begin{tabular}{cccccc}
      \hline\hline
      $\beta$ & Size & $\kappa_{\rm{sea}}$ & $M_\pi$ [GeV] & $L$ [fm] & $M_\pi L$ \\
      \hline
      5.25 & $16^3\times32$ & 0.13460 & 1.281 & 1.22 & 7.9\\ 
      5.25 & $24^3\times48$ & 0.13575 & 0.992 & 1.82 & 9.2\\ 
      5.25 & $24^3\times48$ & 0.13600 & 0.664 & 1.82 & 6.1\\ 
      5.29 & $24^3\times48$ & 0.13550 & 0.896 & 1.73 & 7.8\\ 
      5.29 & $24^3\times48$ & 0.13590 & 0.656 & 1.73 & 5.7\\ 
      5.29 & $24^3\times48$ & 0.13620 & 0.425 & 1.73 & 3.7\\ 
      5.29 & $32^3\times64$ & 0.13632 & 0.295 & 2.30 & 3.4\\ 
      5.40 & $24^3\times48$ & 0.13560 & 1.027 & 1.44 & 7.5\\ 
      5.40 & $24^3\times48$ & 0.13610 & 0.726 & 1.44 & 5.3\\ 
      5.40 & $24^3\times48$ & 0.13640 & 0.506 & 1.44 & 3.7\\ 
      5.40 & $32^3\times64$ & 0.13640 & 0.495 & 1.92 & 4.8\\ 
      5.40 & $32^3\times64$ & 0.13660 & 0.285 & 1.92 & 2.8\\
      \hline\hline
    \end{tabular}
  \end{center}
\end{table}

\section{Methodology}
\begin{figure}
  \begin{center}
    \begin{minipage}{0.45\linewidth}
      \includegraphics[width=\linewidth]{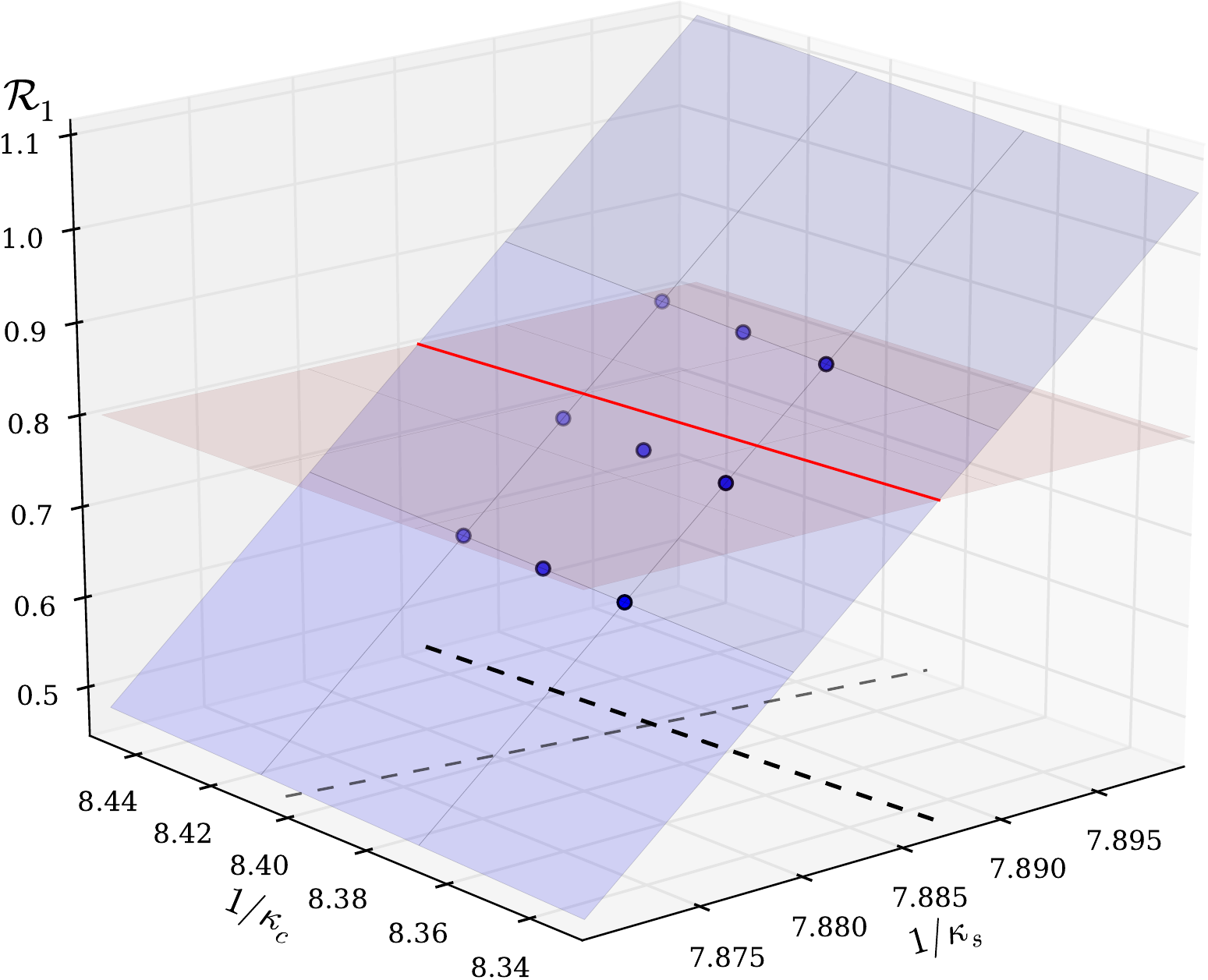}
    \end{minipage}
    \hfill
    \begin{minipage}{0.45\linewidth}
      \includegraphics[width=\linewidth]{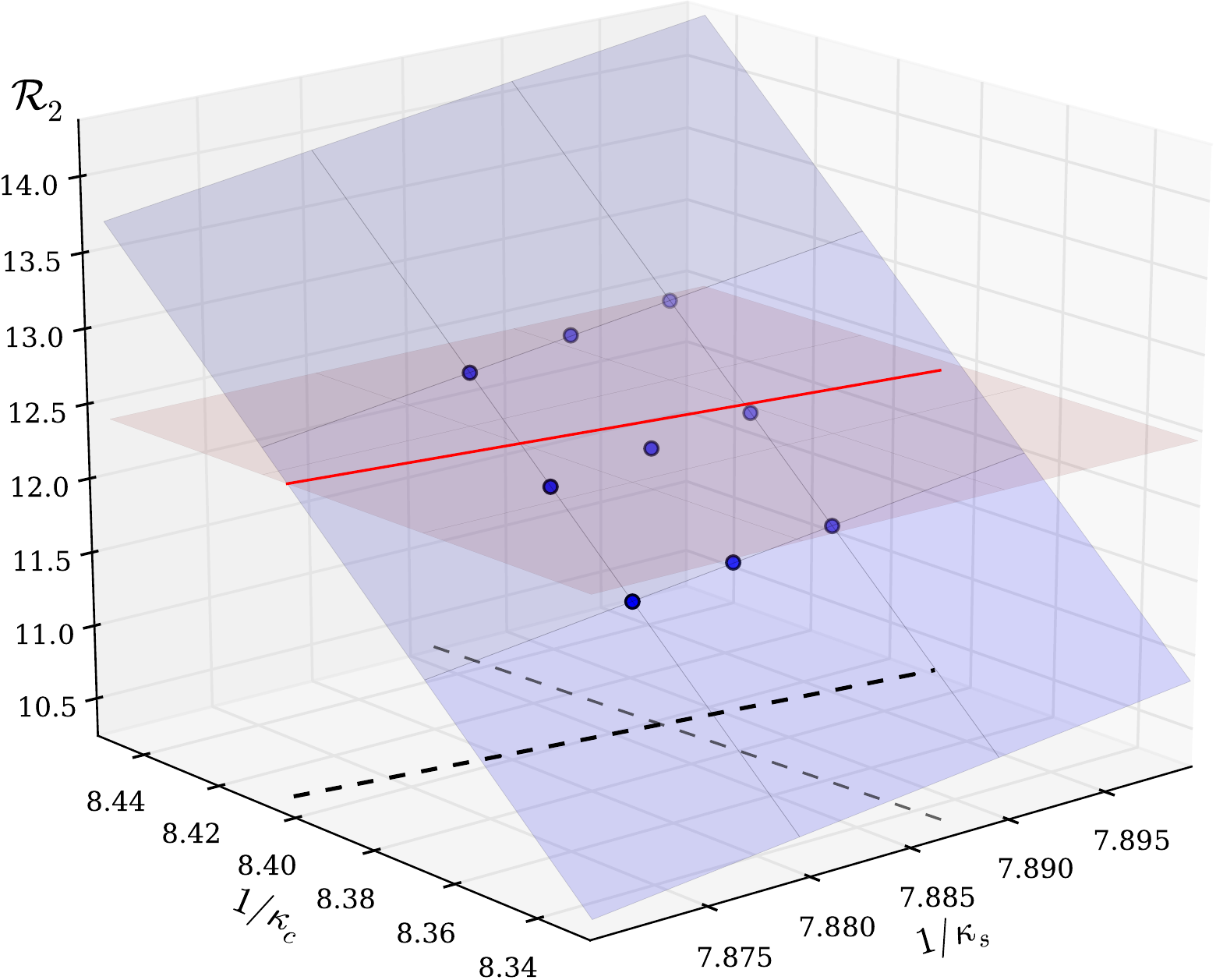}
    \end{minipage}
  \end{center}
  \caption{Example of tuning for $\kappa_s$ and $\kappa_c$. We invert for three values of $\kappa_s$ and $\kappa_c$ giving us nine $R_1$ (left) and $R_2$ (right) ratios (blue circles). The blue surfaces are fits to these nine points. The red surface shows the constant surface of the physical value of the respective ratio, while the red curve is the intersection of the red and blue surfaces (shown as a dashed line on the base-surface of each plot). The point where the two curves intersect is taken as the tuned $\kappa$-pair for this ensemble.}
  \label{fig:r1_and_r2}
\end{figure}

Our mixed action approach requires us to tune the Wilson hopping parameter for the strange and charm quarks. For this tuning we choose ratios of meson masses, which allows tuning these quantities without the need to set the scale. Namely, we choose the two ratios:
\begin{equation}
  \begin{array}{l@{\qquad{\rm and }\qquad}r}
    {R}_1 = \frac{M_{P,\bar{s}s}^2}{M_{V,\bar{c}s}^2-M_{P,\bar{c}s}^2} &
    {R}_2 = \frac{2M_{P,\bar{c}s}^2-M_{P,\bar{s}s}^2}{M_{V,\bar{c}s}^2-M_{P,\bar{c}s}^2}.
  \end{array}
\end{equation}
Note that both ratios depend on both the strange and the charm quark mass. The procedure we follow is to invert for a few choices of the two hopping parameters $\kappa_s$ and $\kappa_c$, and to compute $R_1$ and $R_2$ for every combination of these two. Typically we invert for three $\kappa_s$ and three $\kappa_c$ values, which leaves us with nine points. We then fit to find two curves in the $(\kappa_s, \kappa_c)$ coordinate system along which $R_1$ and $R_2$ acquire their physical values. We take these as $R^*_1$ = 0.801 and $R^*_2 = 12.402$. This gives us a single intersection point, and the $\kappa$-values at this point we call the ``tuned'' kappa-values: $\kappa_s^*$ and $\kappa_c^*$. An example of this tuning procedure for one of the ensembles considered is shown in the plots in Fig.~\ref{fig:r1_and_r2}.

For this calculation, we use the Partially Conserved Axial Current (PCAC) relation:
\begin{align}
      m_{ij}^{PCAC} \equiv \frac{m^{PCAC}_i + m^{PCAC}_j}{2} = \frac{\sum_x\langle\bar{\partial}_t [A_t(x) + a c_A\bar{\partial}_t P(x)]P(0)\rangle}{2\sum_x\langle P(x) P(0)\rangle}
      \label{eq:pcac}
\end{align}
where $i$, $j$ = $c$ or $s$. 
The ratio of quark masses is then given by the ratio of Axial Ward Identity quark masses which are associated with the PCAC quark masses through:
\begin{align}
\frac{m^{AWI}_i}{m^{AWI}_j} = \frac{m^{PCAC}_i}{m^{PCAC}_j}[1+a(m_i^W-m_j^W)(b_A-b_P)+\mathcal{O}(a^2)].
\end{align}
At tree-level, we have $c_A = 0$ and $b_A-b_P = 0$, meaning we can take directly the quark mass ratio $r = m_c/m_s$ from the ratio of PCAC quark masses. A more detailed discussion of these terms is given in Ref.~\cite{Durr:2011ed}. Additionally, we avoid involving correlators of two charm propagators by using the relation:
\begin{align}
      r = \frac{m_c}{m_s} = \frac{2m^{PCAC}_{cs}-m^{PCAC}_{ss}}{m^{PCAC}_{ss}}.
\end{align}
This way the $m_{cc}^{PCAC}$ mass never appears in our analysis. To summarize, the procedure we follow is to first determine the ``tuned'' $\kappa$-pairs as we have already detailed above. Subsequently, we compute the ratio $r$ for each of the nine $(\kappa_s, \kappa_c)$ pairs, as in the case of $R_1$ and $R_2$, and by fitting to a surface we can interpolate for the value of the ratio $r$ at the tuned point $(\kappa^*_s, \kappa^*_c)$. This is taken as the value of $m_c/m_s$ for the given ensemble.

\begin{figure}[!h]
  \begin{minipage}[t]{0.45\linewidth}
    \begin{center}
      \includegraphics[width=1\linewidth]{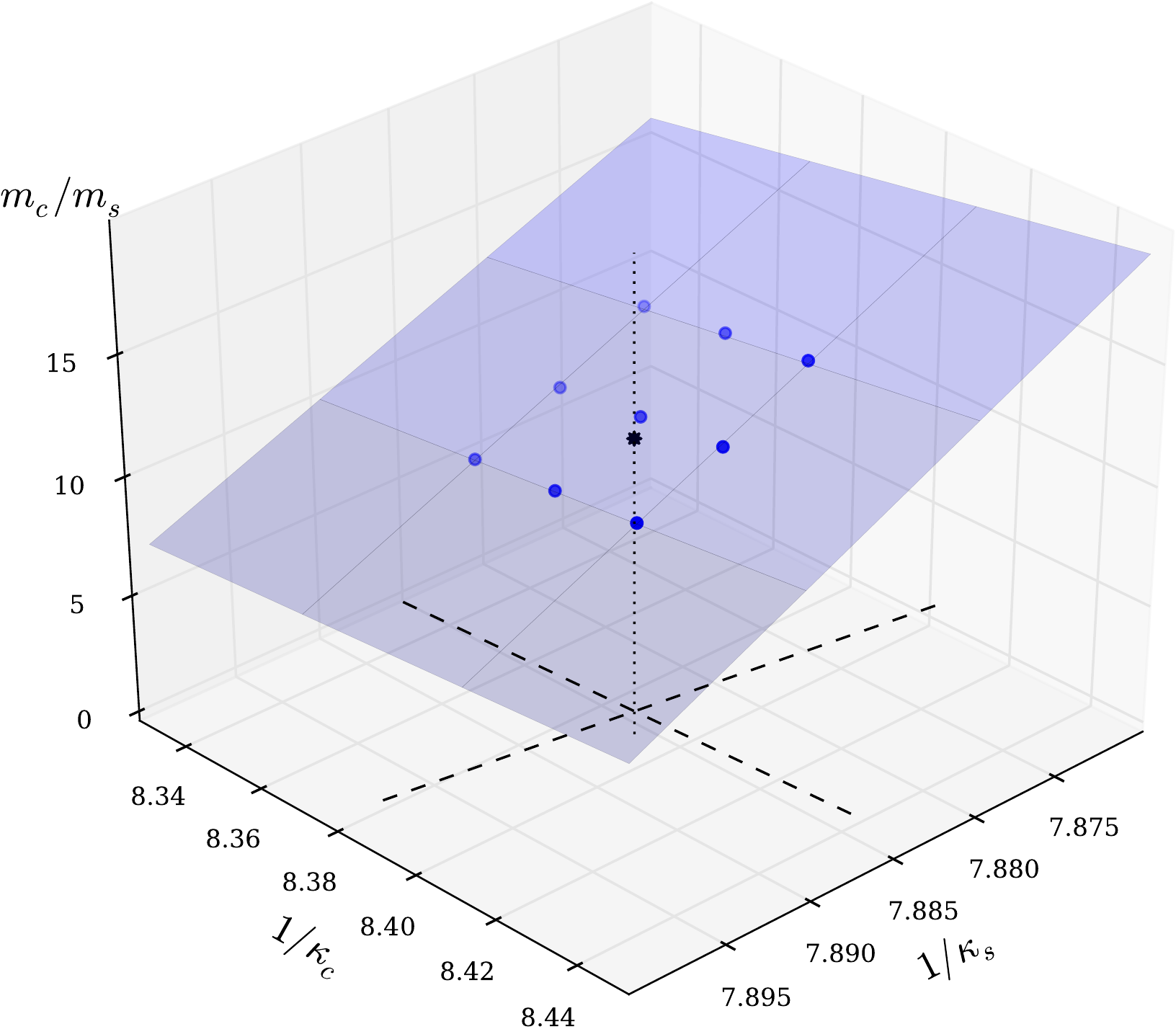}
      \caption{Example of the determination of $m_c/m_s$ at the tuned $\kappa$ values. The blue surface is a fit to the blue circles. The dashed lines cross at the tuned $\kappa$ values, as shown in Fig.~{\protect \ref{fig:r1_and_r2}}. The black asterisk is the central value of $m_c/m_s$ for this ensemble. This is done for each Jack-Knife sample to obtain a Jack-Knife error for the ratio.}
    \label{fig:mcoms_3d}
    \end{center}
  \end{minipage}
  \hfill
  \begin{minipage}[t]{0.45\linewidth}
    \begin{center}
      \includegraphics[width=1\linewidth]{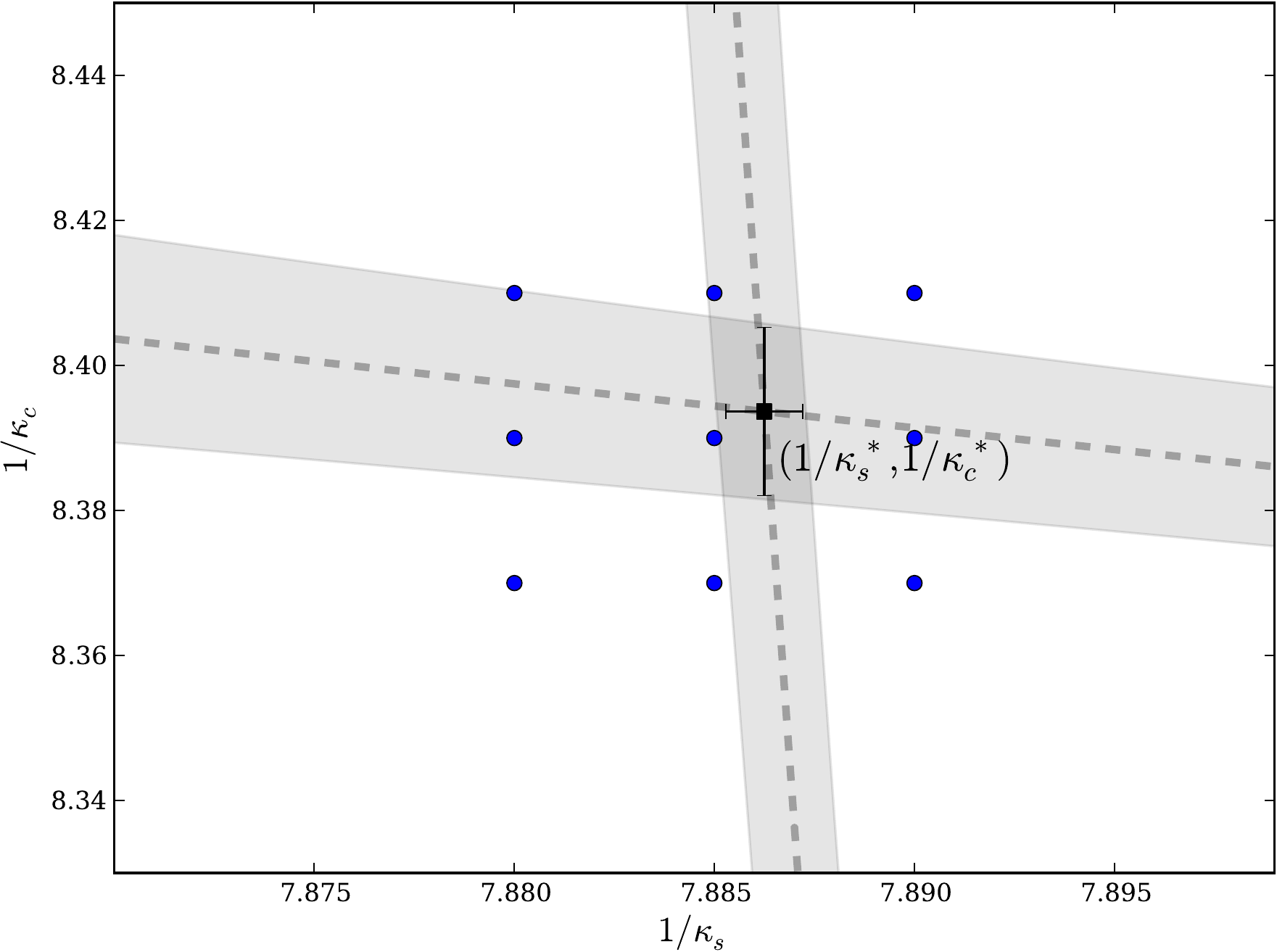}
      \caption{Example of the error associated with the determination of the tuned $\kappa$ values. The plot is a top-down view of Fig.~{\protect \ref{fig:mcoms_3d}}, with error bands added to the dashed lines. The error bands where obtained through a Jack-Knife analysis.}
    \label{fig:crossing}
    \end{center}
  \end{minipage}
\end{figure}
This entire procedure is carried out within a Jack-Knife analysis in order to propagate the statistical error to the final value of $m_c/m_s$. In Fig.~\ref{fig:mcoms_3d} we show an example of the interpolation to the tuned $\kappa$-values. In Fig.~\ref{fig:crossing}, we show the statistical errors associated in determining the tuned $\kappa$-values. These are obtained by performing a Jack-Knife analysis, where the crossing is determined in every Jack-Knifed sample. We note that the errors on $\kappa_s^*$ and $\kappa_c^*$ are not needed in the final analysis, since the Jack-Knife procedure is carried out up to the final determination of $r$. The errors in Fig.~\ref{fig:crossing} have been extracted for illustration purposes only.

\section{Results}
\begin{figure}
  \begin{minipage}[t]{0.32\linewidth}
    \includegraphics[width=\linewidth]{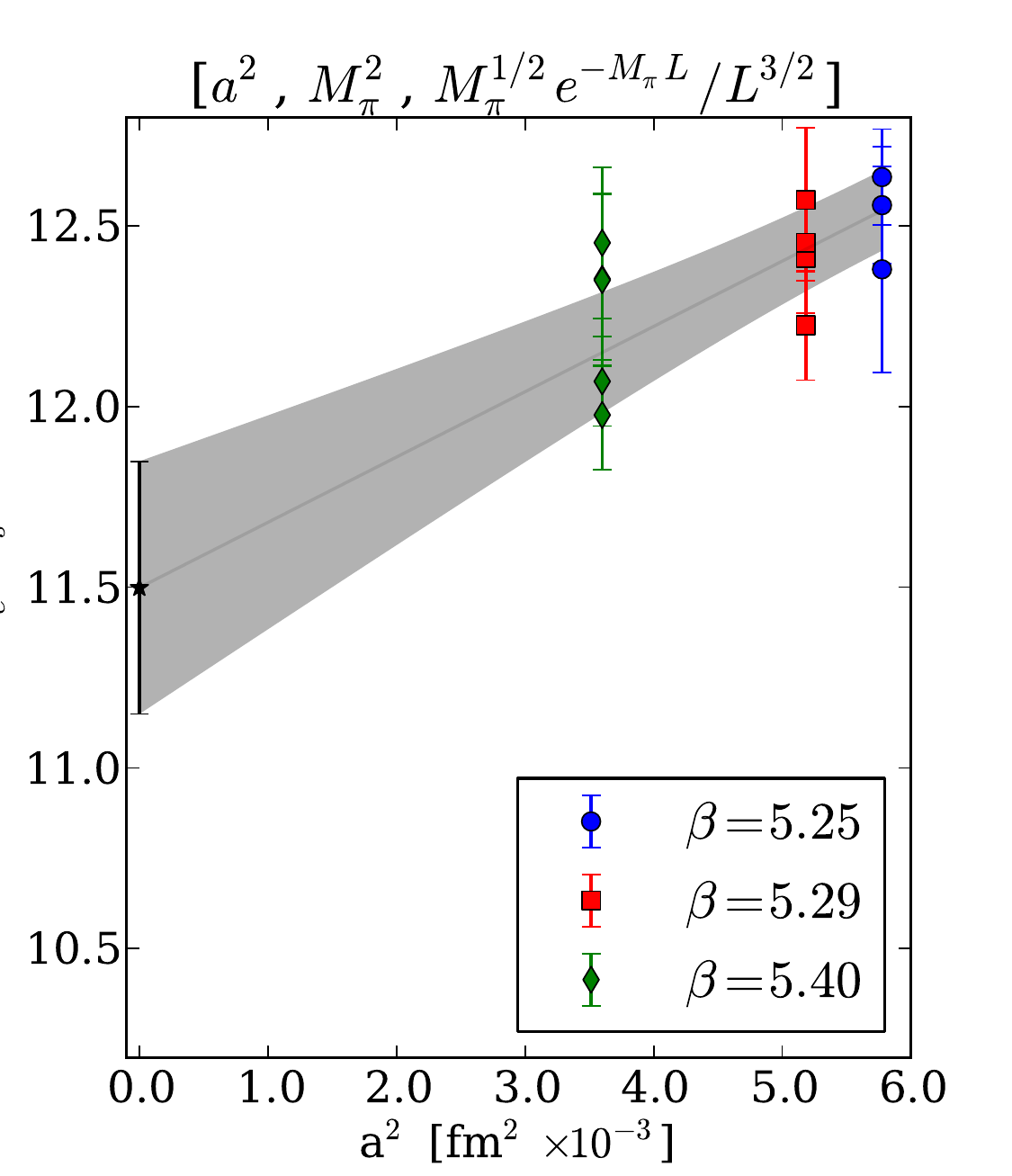}
  \end{minipage}
  \hfill
  \begin{minipage}[t]{0.32\linewidth}
    \includegraphics[width=\linewidth]{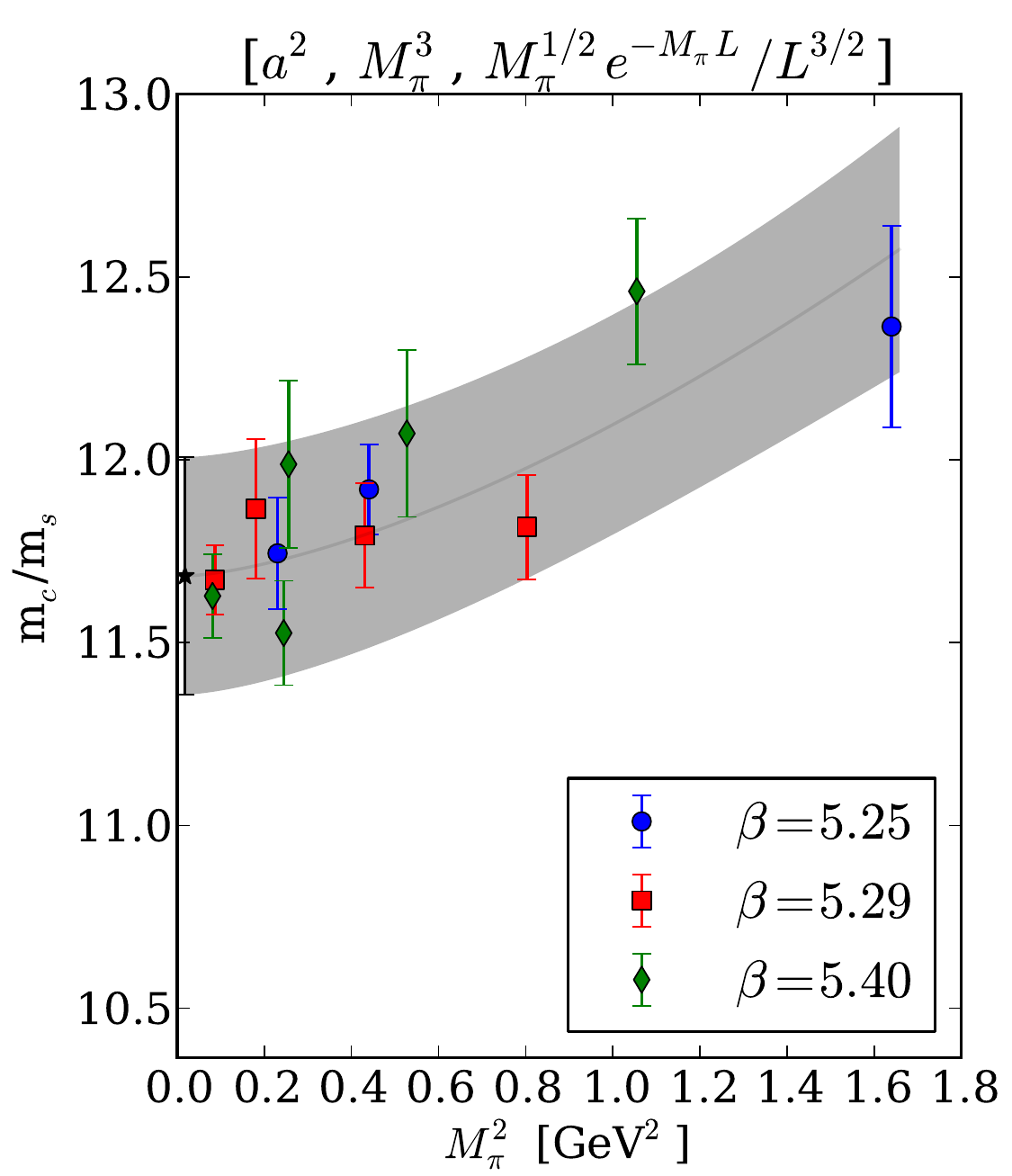}
  \end{minipage}
  \hfill
  \begin{minipage}[t]{0.32\linewidth}
    \includegraphics[width=\linewidth]{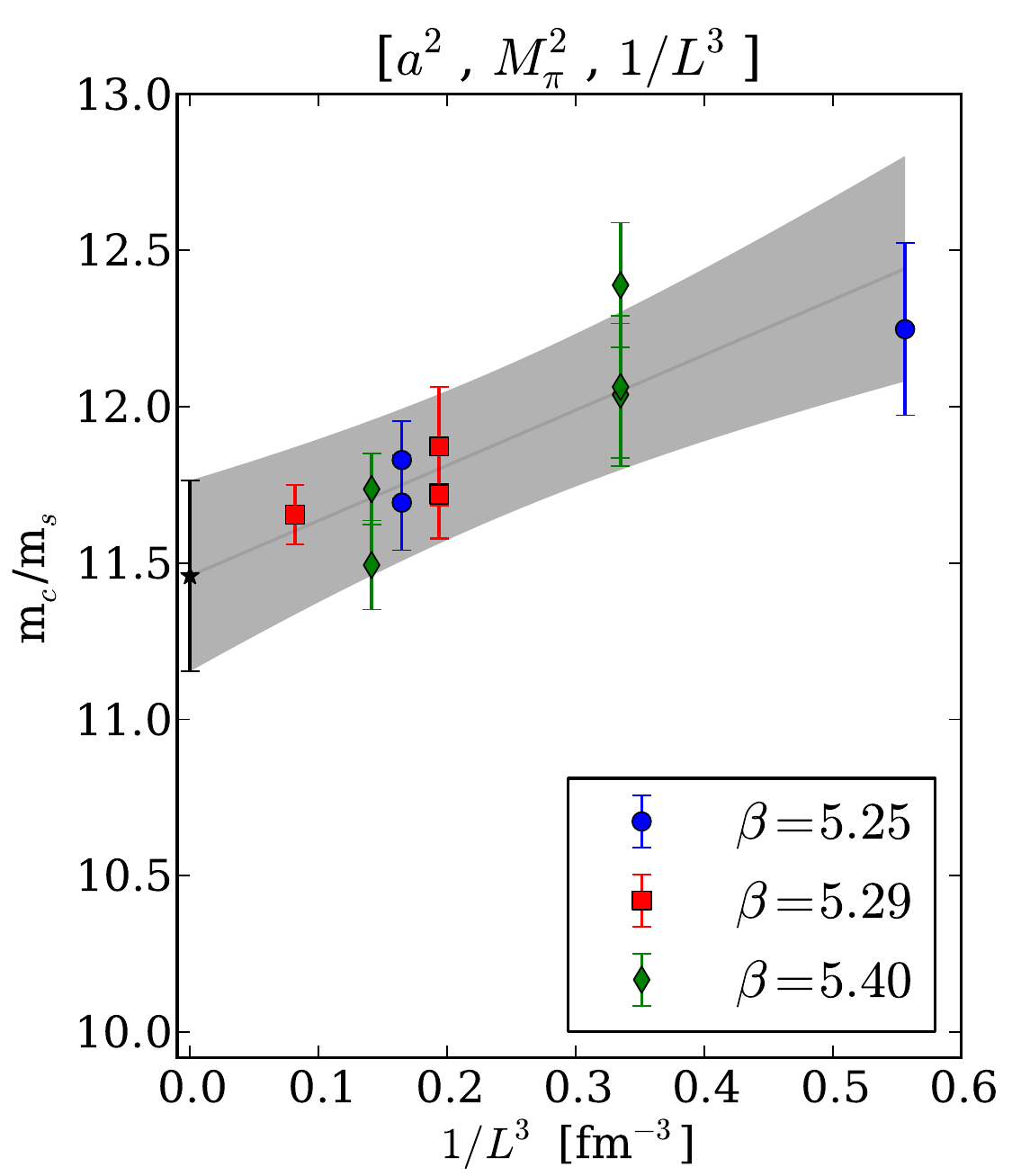}
  \end{minipage}
  \label{fig:fits}
  \caption{Examples of fits of our data to the global fit function Eq.~{\protect \ref{eq:global_fit_func}}. As the fit is a function of three independent variables, in each of these three figures, the data points are extrapolated to the physical values of the two axes not shown. The left plot shows a fit to $f^{1, 0, 0}(a, M_\pi, L)$ as a function of $a^2$, the center plot a fit to $f^{1, 1, 0}(a, M_\pi, L)$ as a function of the pion mass squared and the right plot a fit to $f^{1, 0, 1}(a, M_\pi, L)$ as a function of the inverse volume.}
\end{figure}

As mentioned, the ensembles were chosen in order to allow for a reliable extrapolation to the physical pion mass, infinite volume and zero lattice spacing. We perform a global fit so as to extrapolate simultaneously to these three physical limits. For each of the three extrapolations we take two fit ans\"atze which means these can be combined in eight ways, thus yielding eight global fit functions:
\begin{equation}
  f^{i,j,k}(a, M_\pi, L) = c_0\left[1 + c_1^i F^i(a) + c_2^j G^j(M_\pi) + c_3^k H^k(M_\pi, L)\right]\quad\textrm{with}\quad i,j,k = 0\textrm{ or }1
  \label{eq:global_fit_func}
\end{equation}
where:
\begin{equation}
  \begin{array}{rl@{\qquad}rl}
    F^0(a) &= \alpha_s a & F^1(a) &= a^2 \\
    G^0(M_\pi) &= M_\pi^2 & G^1(M_\pi) &= M_\pi^3 \\
    H^0(M_\pi, L) &= \sqrt{\frac{M_\pi}{L^3}} e^{-M_\pi L} & M^1(M_\pi, L) &= \frac{1}{L^3}.
  \end{array}
\end{equation}
In Fig.~\ref{fig:fits} we show three plots of representative extrapolations. In each plot, the extrapolation is shown with respect to one of the three extrapolation axes. For illustration purposes, the data in the plots have already been extrapolated for the axes not shown.

We note here that we considered two possible ways to discretize the partial derivative involved in the calculation of the PCAC quark mass (Eq.~\ref{eq:pcac}). Namely, we considered the regular symmetric derivative: $\bar{\partial}_t f(t) = [f(t+1)-f(t-1)]/2$ as well as the improved form:
\begin{align}
  \bar{\partial}_t f(t) = \frac{f(t-2)-8f(t-1)+8f(t+1)-f(t+2)}{12}.
\end{align}
Although the two forms yield consistent results for $m_c/m_s$, using the latter form we obtain $\chi^2/\rm d.o.f$ of order one for each of the eight extrapolations, while using the regular form we obtain $\chi^2/\rm d.o.f$ in tendency above one. We therefore opt to use the improved form as the definition of the discretized derivative $\bar{\partial}_t$ throughout our analysis.

\begin{figure}
  \begin{minipage}[t]{0.45\linewidth}
    \begin{center}
      \includegraphics[width=1\linewidth]{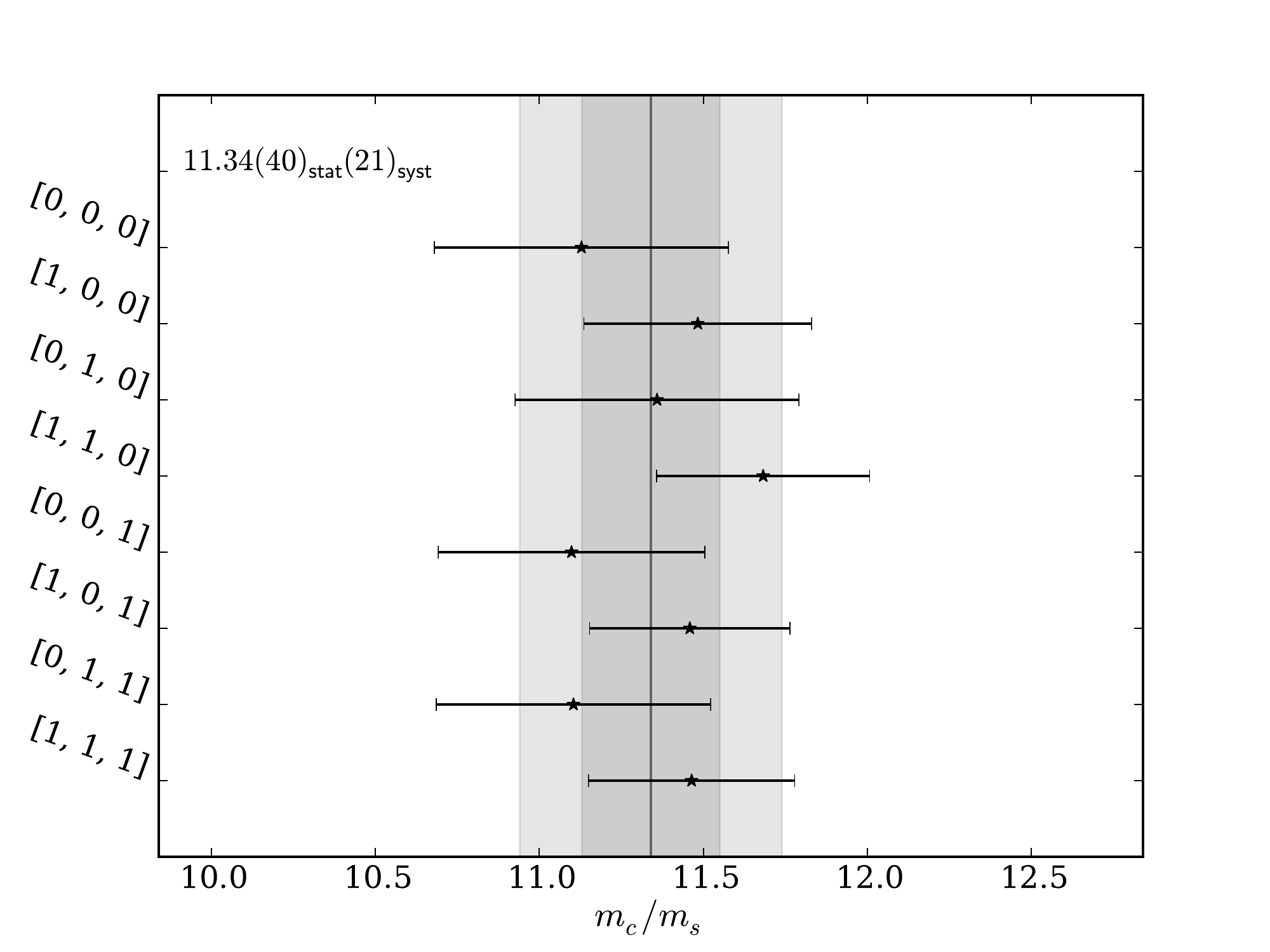}
      \caption{The ratio $m_c/m_s$ as obtained from the eight fit functions. The solid vertical line is the average of the eight values. The lighter gray band indicates the statistical error of the central value while the darker gray band shows the systematic error, obtained from the standard deviation of the eight values.}
    \label{fig:mcoms_result}
    \end{center}
  \end{minipage}
  \hfill
  \begin{minipage}[t]{0.45\linewidth}
    \begin{center}
      \includegraphics[width=1\linewidth]{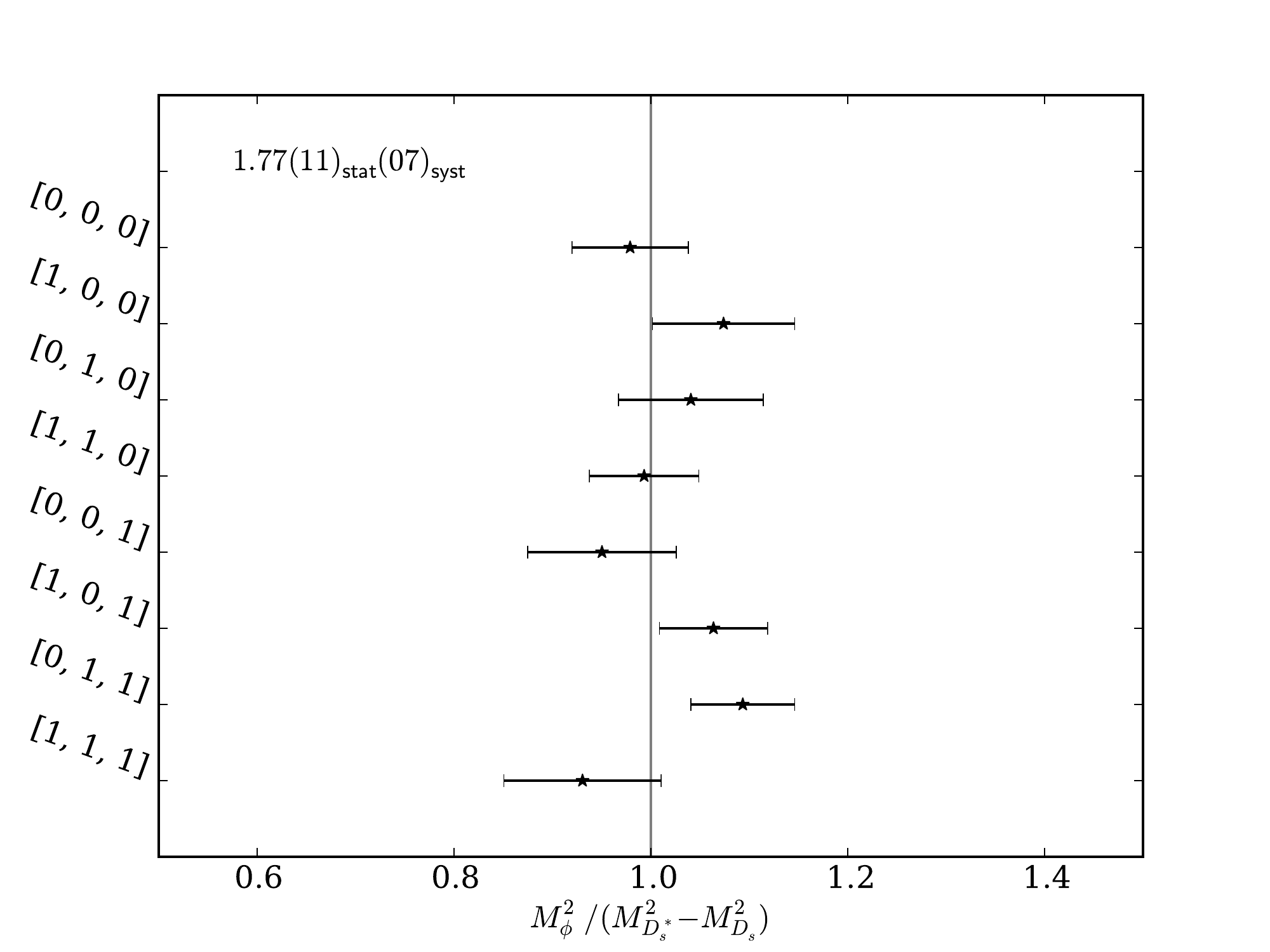}
      \caption{The ``control ratio'' $R_3$ obtained by the eight fits, normalized by its physical value of 1.7707.}
      \label{fig:control}
    \end{center}
  \end{minipage}
\end{figure}

The fit procedure described gives eight values of the ratio $m_c/m_s$ at zero lattice spacing, infinite volume and the physical pion mass, one for each of the eight $[i, j, k]$ combinations of Eq.~\ref{eq:global_fit_func}. We combine these eight values to obtain our final estimate of the ratio of $m_c/m_s$, by taking the standard deviation of these eight values as the systematic error of our measurement, while the average of the eight central values and of their statistical error is taken as the central value and statistical error respectively. The eight results of the fit are shown in Fig.~\ref{fig:mcoms_result} with the error-bands denoting the systematic and the statistical error. Our final value for the ratio of $m_c/m_s$ is:
\begin{equation}
  m_c / m_s = 11.34(40)_{\rm stat}(21)_{\rm syst}.
\end{equation}
As a potential check of our method, we carry out an identical analysis for the ratio $R_3 = \frac{M_\phi^2}{(M_{D_s^*}^2 - M_{D_s}^2)}$, of which the physical value is well determined at $1.7707$. The result of the eight fits for this ``control quantity'' is shown in Fig.~\ref{fig:control} (normalized to the physical value). The result is $R_3 = 1.77(11)_{\rm stat}(07)_{\rm syst}$. The excellent agreement of this quantity with its physical value indicates that our tuning procedure and analysis are sound and that the obtained estimate of $m_c / m_s$ is reliable.

\section{Summary}

In this presentation we have carried out a determination of the fundamentally interesting quantity of the charm to strange quark mass ratio $m_c/m_s$. This was done using a relativistic fermion action in the valence sector with improved scaling at large quark masses. We perform a controlled extrapolation to the physical pion mass, zero lattice spacing and infinite volume.

\begin{figure}[!h]
  \begin{center}
    \includegraphics[width=\linewidth]{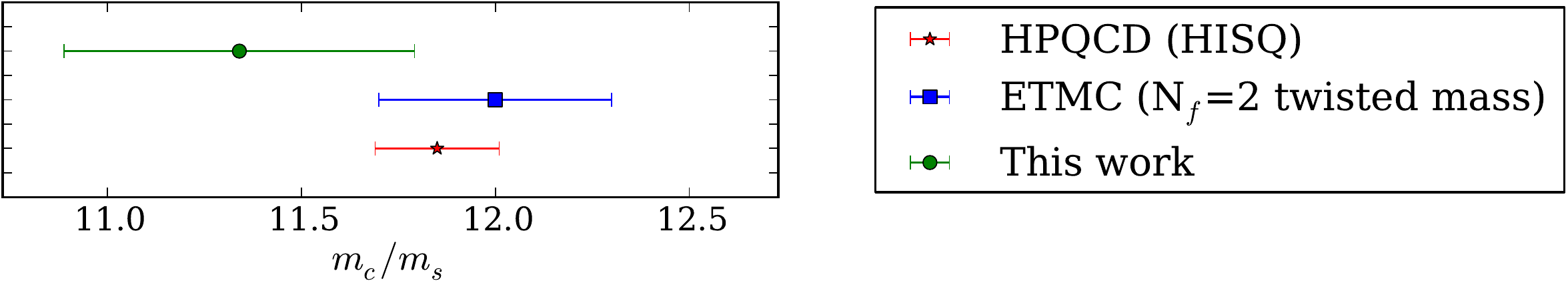}
  \end{center}
  \caption{Three results of the ratio $m_c/m_s$ computed on the lattice. The red star corresponds to the result by the HPQCD collaboration, the blue square is the result by the ETM collaboration and the green circle is the result of this work.}
  \label{fig:mcoms_comparison}  
\end{figure}

Our result is in reasonable agreement with two other recent calculations of this quantity. Namely, the first result obtained using a relativistic fermion action was a calculation using staggered quarks~\cite{davies:2009ih} which arrived at $m_c/m_s = 11.85(16)$, while a second result available by the ETM collaboration~\cite{blossier:2010cr} arrived at $m_c/m_s = 12.0(3)$, where the latter's error is statistical with no clear indication of the systematic uncertainty involved. A comparison of our result to these two determinations is shown in Fig.~\ref{fig:mcoms_comparison}. Although our result is somewhat more imprecise, it serves as an important check since our fermion action is free from any isospin or taste symmetry breaking which is inherent in either staggered or twisted mass fermions. 

Additionally, we note that the computer resources required for this calculation where rather modest. The improved approach to the continuum exhibited by the Brillouin improved fermion action allowed a reliable extrapolation to the physical point with a moderate number of statistics and gauge ensembles. Namely, the requirements for this calculation where roughly 20 node-years, of an eight-core Nehalem node. This number includes inverting for three kappa values for each of the charm and strange quark masses, as well as trial inversions at each ensemble to determine the best three such values.

\bigskip
\noindent\textbf{Acknowledgments:} We thank the QCDSF collaboration for allowing us to use their N$_{\rm f}$ = 2 configurations \mcite{
Bietenholz:2010az, Collins:2011mk} and the ILDG for making them available. We would like to thank Stefan Sint for useful feedback. We acknowledge partial support in SFB/TR-55. CPU resources on JUROPA were provided by Forschungszentrum J\"ulich GmbH through a VSR grant.

\bibliographystyle{JHEP}
\bibliography{refs}

\end{document}